\documentclass{aastex}
\usepackage{emulateapj5,apjfonts,psfig}

\def\asca{{\sl ASCA }}

\def\ros{{\sl ROSAT }}

\def\ein{{\sl Einstein }}

\def\chandra{{\sl Chandra }}

\def\ergsec{\hbox{erg s$^{-1}$ }}
\def\ergcm{\hbox{erg cm$^{-2}$ s$^{-1}$ }}

\def\Msun{$M_{\odot}$ }

\def\it{\sl}

\slugcomment{accepted for publication in The Astrophysical Main Journal}

\shorttitle{RESOLVING THE ORION TRAPEZIUM IN X-RAYS}

\shortauthors{SCHULZ et al.}

\begin{document}

\title{Chandra Observations of Variable Embedded X-ray sources in Orion. \\
Paper I: Resolving the Orion Trapezium}
\author{
N.S. Schulz\altaffilmark{1},
C.Canizares\altaffilmark{1},
D.Huenemoerder\altaffilmark{1},
J.H.Kastner\altaffilmark{2},
S.C.Taylor\altaffilmark{1},
E.J.Bergstrom\altaffilmark{2}
}
\altaffiltext{1} {Center for Space Research, Massachusetts Institute of Technology, Cambridge MA 02139, USA}
\altaffiltext{2} {Chester F. Carlson Center for Imaging
Science, Rochester Institute of Technology, Rochester NY 14623, USA}

\begin{abstract}
We used the
High Energy Transmission Grating Spectrometer (HETGS) onboard
the Chandra X-ray Observatory to perform two observations, 
separated by three weeks, of the Orion Trapezium region. 
The zeroth order images on the Advanced CCD Imaging Spectrometer (ACIS)
provide spatial resolution of 0.5" and moderate energy resolution.
Within a 160"x140" region around the Orion Trapezium we resolve 111 X-ray sources
with luminosities
between 7$\times 10^{28}$ \ergsec and 2$\times 10^{32}$ \ergsec. We do not detect any 
diffuse emission. All but six sources are identified.
From spectral fits of the three brightest stars in the Trapezium we
determine the line of sight column density to be N$_{H}$ = 1.93$\pm0.29 \times 10^{21}$ cm$^{-2}$. 
Many sources
appear much more heavily absorbed, with N$_{H}$ in the range of $10^{22}$ to $10^{23}$ cm$^{-2}$.
A large fraction of sources also 
show excursions in luminosity by more than a factor 5 
on timescales $>$50 ks; many are only detected in one of the observations.

The main objective of this paper is to study the Orion Trapezium and its close vicinity.
All five Trapezium stars are bright in X-rays, with $\theta^1$ Ori C accounting for about
60$\%$ of the total luminosity of the Trapezium.
The CCD spectra of the three very early type members can be fit with a 
two-temperature thermal spectrum with a soft component of kT $\sim$ 0.8 keV and a hard component of 
kT $\sim$ 2 to 3 keV.
$\theta^1$ Ori B is an order of magnitude fainter than
$\theta^1$ Ori E and shows only a hard spectrum of kT $\sim$ 3 keV. 
$\theta^1$ Ori D is another order of magnitude fainter than $\theta^1$ Ori B, with only a  kT $\sim$ 0.7 keV component.
We discuss these results in the context of stellar wind models.

We detect eight additional, mostly variable X-ray sources in the close vicinity of the Trapezium.
They are identified with thermal and non-thermal radio sources, as well as 
infrared and optical stars. Five of these X-rays sources are identified with proplyds   
and we argue that the X-ray emission originates from class I, II and III protostars at
the cores of the proplyds.

\end{abstract}

\keywords{
stars: clusters ---
stars: formation ---
stars: early type ---
X-rays: stars ---
techniques: spectroscopic}

\section{Introduction}

The Trapezium Cluster of the Orion Nebula is one of the best
studied star forming regions. 
At a distance of
440 pc, the Trapezium Cluster is embedded at the near rim of the giant dark molecular
cloud L1641, lying just behind the Orion
Nebula (NGC 1976). It is the core of the Orion Nebula Cluster (ONC), a 
loose association of more than  5000 mostly very young ($<$ 1-2 Myr) 
pre-main sequence stars of a wide range of stellar mass ($\sim 0.1 - 50$ \Msun)
within a radius of about 1.3 pc (see Hillenbrand 1997 and references therein). A density of $\sim 5 \times 10^3$ stars
per pc$^3$ within the central 0.1 pc makes the Trapezium Cluster the most dense stellar 
clustering known in our Galaxy (McCaughrean $\&$ Stauffer 1994) with about 3000 \Msun pc$^{-3}$.

The Orion Trapezium was discovered by Trumpler (1931) and its main
stellar component, $\theta^1$ Ori C, was identified as the main source of excitation of the Orion
Nebula. 
Many observations
of the Trapezium have been made at infrared wavelengths, mainly the K- and H-bands
at 2.20 $\mu$m and 1.65 $\mu$m respectively. 
$\theta^1$ Ori A  - E are the brightest 
stars in the cluster, optically and at infrared
wavelengths, and  are 
early type  O or B-stars.
Three members of the Trapezium are multiple systems,
some with separations much smaller than
90 AU (Petr et al. 1986). 
About 80 $\%$ of the stars are younger than 1 Myr with
a median age of 3$\times 10^5 $ yr. The present star formation rate
is estimated at  
$\sim 10^{-4}$ \Msun yr$^{-1}$ (Hillenbrand 1997).

The core of the Orion Nebula is one of the prime areas to observe and study very 
young stellar objects (YSOs). Laques $\&$ Vidal (1979) detected externally illuminated   
YSOs in the form of compact H$\alpha$ brightness knots, which are illuminated by the two optically brightest
sources, 
$\theta^1$ Ori C and $\theta^1$ Ori A. Garay et al. 1987 measured free-free
radio continua from most of these sources. 
Many so-called 'proplyd' (externally illuminated protoplanetary disk)
candidates have been discovered in the vicinity of the
Trapezium (O'Dell et al. 1993), many of which correspond to
radio and/or IR sources (Felli et al. 1993; McCaughrean and Stauffer 1994).
Bally et al. (1998) studied 
over 40 proplyds with HST and discussed
models for these illuminated YSOs involving a young star at the core with a circumstellar disk
embedded in an outer wind shock region, as well as a wind driven tail pointing away from the
illuminating star.    

Because of its proximity, the Orion Trapezium came under early
scrunity in X-rays and it was UHURU, the first X-ray satellite 
(Giacconi et al. 1971), that discovered X-rays from that region in 1972. 
The \ein Observatory for the first time observed
X-rays identified with main sequence O6 to B5 stars (Cassinelli et al. 1981). These early type stars do not have a
stellar corona like late type stars, but exhibit a very hot stellar wind that
ionizes its environment. Here X-rays are generally thought to originate from
bow shocks surrounding radiatively driven blobs plowing through rarified
wind material (Lucy and White 1980). Surprisingly, X-rays from some late-B and early
A-type stars were discovered, where neither a corona nor an important stellar wind
is thought to exist. The generally favored explanantion for these X-rays is that they
originate not from the star itself but from an unresolved, otherwise invisible
low-mass companion (Bergh\"ofer and Schmitt 1994). On the other hand the ONC hosts a large sample
of late type pre-main sequence (PMS) stars. \ein and \ros established a large catalog of X-ray sources 
within an area of 5'x5', of which over 200 objects were identified as PMS stars with spectral types of F
to late G (Ku $\&$ Chanan 1979, Gagne et al. 1995, Geier et
al. 1995).
These stars generate X-rays through magnetic activity and a hot corona.  
It has also been suggested 
that the Trapezium Cluster 
itself features a faint diffuse X-ray component from ionized material in between
the Trapezium stars (Ku $\&$ Chanan 1979). 

With the launch of ROSAT in 1990, individual X-ray sources were resolved in the Trapezium
(Caillault et al.  1994; Gagne et al. 1995), however the question of a diffuse X-ray component
within the Trapezium remained (Geier et al. 1995).
The \asca observatory launched in 1993 observed the Orion Trapezium cluster with its
until then unprecedented spectral resolving capacity up to 10 keV. 
The results showed that a high temperature plasma
with kT $\sim 2-5$ keV exists in the Trapezium, and 2-temperature thin thermal
plasmas are more likely than single temperatures (Yamauchi et al. 1996). This is in agreement with the
observations of O-stars with \asca (Corcoran et al. 1994), which obtained spectra of $\delta$ Ori
and $\lambda$ Ori characterized by two-temperature models with additional absorption attributed to a warm wind. The
hot components had temperatures of kT $\sim$ 0.6 and 2 keV. The \asca observations also did not observe
the large flares one expects from late type T Tauri stars, but the spectra showed
the existence of a variable hard spectral component associated with the Trapezium.

The physical origin of highly energetic emission from
young stars is still uncertain.
So far only empirical relationships exist between this emission and stellar age, size and mass
of the star, as well as stellar rotation (Feigelson et al. 1993,
Kastner et al. 1997).
We currently classify evolutionary traces simply by observability in the radio to X-ray band.
Class 0 sources are considered to be still deeply embedded and merely gravitationally collapsing
cores and the spectral energy distribution peaks in the far-IR/submm range.
Class I sources are accreting from a protostellar cloud and are optically obscured. The slope of these
spectra is known to rise from near-IR through far-IR.
Class II sources or classical T Tauri stars (CTTS) are strongly visible in the optical in the H$\alpha$-line
but also have a substantial IR excess. Class III sources or weak line T Tauri stars (WTTS) are
optically bright with little or no IR excess; see
the review by Feigelson $\&$ Montmerle (1999) and references therein.
\asca recently detected giant X-ray flares with variable hard X-ray emission from YSOs (Kamata et al. 1997,
Tsuboi et al. 1998, Yamauchi $\&$ Kamimura 1999, Tsuboi et al. 2000). Spectral temperatures ranged from
kT $\sim$ 4 - 6 keV, in extreme cases up to $>$ 10 keV. A recent modeling of these X-ray flares from
two class I protostars relates this emission to strong magnetic shearing and reconnection between the
central star and the sourrounding protostellar accretion disk (Montmerle et al. 2000).

The next steps toward a better understanding of the
astrophysical processes underlying high energy emission at
early stages of stellar evolution can
be made with the Chandra X-ray Observatory. The Orion Nebula Cluster
was observed early in the Chandra mission by Garmire et
al. 2000, who detected
about 1000 resolved X-ray sources, the richest field of sources
ever imaged in X-rays. 
The observation was performed with a wide field of
view and therefore was best suitable for studying the overall demographics
of the X-ray emission in the whole Orion Nebula Cluster. From the placement of the
identified sources in the Hertzsprung-Russell (HR) diagram
it was concluded that a
large fraction of the cluster members with masses above 1
\Msun are detected in X-rays, whereas 
the detection rate decreases rapidly for
lower mass stars.

In our observations we focus on a smaller field of view around
the Trapezium, which suffers from saturation effects in
the Garmire et al.\ (2000) data. Our objective is to 
resolve the Orion
Trapezium cluster core in X-rays and to perform a detailed spectral 
analysis on the bright sources as well as to study the variability of the X-ray emission. 
We present the results from the zero order field analysis in three parts. Paper I  
gives an overview of these observations but focuses on the X-ray properties of the
Orion Trapezium and its very close vicinity within a radius
of 15". Paper II (Kastner et al. 2000,
in preparation) then discusses identifications within the whole available field
of view around the Orion Trapezium with known infrared and radio catalogs as well as with
newly obtained IR observations. In paper III (Schulz et al. 2000a, in preparation) we
present X-ray spectroscopy on the whole field of view.  
Schulz et al. (2000) also present high resolution spectroscopy of $\theta^1$ Ori C using
the first order spectra of the HETGS.

\section{Chandra Observations and Data Analysis}

The Chandra X-ray Observatory (Weisskopf et al. 1996) was launched on the 
23rd of July 1999 into  
a highly eccentric 72 h orbit to allow uninterrupted observations of up to 140 ks.
Chandra has an unprecedented spatial resolution of 0.5
arcsec (225 AU at the distance of the Trapezium). 
Besides its 
spatial resolving capabilities, Chandra also allows for medium resolution
spectroscopy ($E/{\Delta}E$=6-60) in the 0.1 to 10.0 keV
band using the Advanced CCD Imaging Spectrometer (ACIS) and
high resolution spectroscopy with $E/{\Delta}E$ of
up to 1200 at 1 keV using gratings.
For detailed descriptions of the spectroscopic
instruments see Garmire 
et al. (2000, in preparation), Markert et al. (1994) and Canizares et al. (2000, in preparation). 

We obtained two observations of the Trapezium Cluster using the High Resolution Transmission
Grating Spectrometer (HETGS), the first on October 31st UT 05:47:21 1999, the second about three
weeks later on November 24th UT 05:37:54 1999, for durations of 50 ks and 33 ks, 
respectively. In this paper we report on data obtained with the CCD camera in the 
0th order of the HETGS with the aimpoint on the Orion Trapezium (RA $5^h35^m14.5^s$ and Dec $-5^o23'32''$),
which was positioned on the middle of the central back-illuminated CCD device S3 on node 0. 
This device was unaffected by the degradation of charge transfer inefficiency
of the ACIS front illuminated CCDs due to high dosage of particle irradiation
early into the mission as reported by Prigozhin et al. (2000).
Because the zeroth order image contains roughly 30$\%$ of the flux that would be incident on ACIS if HETG 
were not inserted, it is less susceptible
to the effects of photon pileup in the CCDs. There
are no apparent image distortions due to pileup, and only the CCD spectrum of the brightest
source, $\theta^1$ Ori C, appears to be affected by pileup.

The Chandra X-ray Center (CXC) provided aspect corrected level 1 event lists via
standard pipeline processing. We reprocessed these data 
using the most updated calibration data products and ACIS-S detector
to sky transformations available to us. We also removed all events that resulted from either bad pixels
or columns as well as from flaring background events that
were not removed
by the CXC standard processing. The aspect solution still contains a systematic
uncertainty of the order of 1-2 arcseconds. We remove these systematics
by applying SIMBAD positions of 3 very bright and identified stars in the 
field of view. With this technique we are able to reduce any systematic offsets
between the 1999 Oct. and the 1999 Nov. observations to less than half a pixel or
0.2 arcsec. All currently available calibration products are normalized to 
a split grade selection of ASCA grades of 0, 2, 3, 4, and 6.   
For the data analysis we mainly used FTOOLS 4.2, XSPEC 10.0, and custom software. Errors
of the spectral analysis are based on $90\%$ confidence. Errors in fluxes and luminosities are based
on the current status of effective area calibration on the
ACIS-S3 device, which predicts uncertainties of the order of 10$\%$ over the
energy range between 0.6 and 8 keV. 

For source detection we use the standard `celldetect' algorithm offered by CXC with a 
signal to noise threshold set to 10 $\sigma$. However since we 
allow a lower detection limit of 7 counts above background within 0.7" radius, we added a few more
fainter sources to the list by hand. The average background level within this selection
radius was about 1 count. This correponds to a sensitivity limit
of $3\times10^{-15}$ \ergcm or a lower limit in luminosity 
of $6.6\times10^{28}$ \ergsec (for these estimates
we applied a 1 keV Raymond-Smith model with the column density determined in section 4). In the second observation
this limit is slightly higher because of the shorter exposure, resulting in a limiting luminosity of
$1.0\times10^{29}$ \ergsec. Because of the likelihood of
source variability we ran source detection
on the two single fields of view as well as the merged fields. The  
sensitivity limit of the latter is 4.0$\times10^{28}$sec. For the search for diffuse emission we are not
bound to the source detection radius of 0.7", but applied the detection threshold to a unit area of 
one arcsec$^2$, which translates into a sensitivity limit of 2.0$\times10^{28}$ \ergsec arcsec$^2$.  

\section{Wide Field Observations}

Figure 1 shows a 160"x160" area around the Trapezium , the October 
view in 1a, the November view 1b. This region is
relatively free of photons dispersed into first and higher
orders by the gratings. 

The colors in Figure 1 represent different energy bands.
Red was selected for 0.1 to 1.0 keV, green for 1.0 to 1.5 keV, and blue for 1.5 to
10.0 keV. The energy boundaries were chosen to represent typical stellar spectra (Raymond-Smith models around 1 keV) for moderate
galactic absorption. In this respect a star that appears red has either a very soft spectrum
or appears in the foreground. 
On the other hand, if a star appears blue it is likely absorbed and
has a hard spectrum; while
if it appears white, it likely has a 
hard but less absorbed spectrum.

A comparison of the two star
fields shows that the whole field is resolved into point sources. We detected 111 sources within a field of
160"x140". These sources are listed in table 1 with
count rates for each of the two observations, as well as
the total number of detected counts in the combined
(1999 Oct. plus 1999 Nov.) data set. The sources are labelled
A-E for the main Trapezium stars and then by numbers and are sorted by RA
(2000). So far we identified all but 6 sources in the field and we show the various references in table 1.
A more thorough discussion of these identifications will follow in paper II. The central star of the Trapezium is also 
the brightest X-ray object in the field. Most sources,
however, are very faint. Only 7 of the
111 detected sources show X-ray fluxes
$>10^{-12}$ \ergcm ($>2\times 10^{31}$ \ergsec).
Many sources in Fig.\ 1  appear blueish
in color. As a guideline, the central star
$\theta^1$ Ori C has the assumed column density (see Sec. 4) and
appears more white and red in color in Fig.\ 1. 

Clearly
these blue sources are significantly more absorbed than $\theta^1$ Ori C, which indicates that they
either are intrinsically absorbed or lie deeply embedded in L1641.
This is consistent with the earlier
observation by Garmire et al. (2000). 
Sources in the northeastern
part of the field appear more affected by this, which may indicate that many of them
lie indeed embedded in L1641. 
There is also an indication that most X-ray sources are confined to within an 0.1 pc radius of the Trapezium, i.e   
the source density drops considerably to the south and west of the cluster.

A comparison of figures 1a and b
show that most of the absorbed blue
sources also appear to be the most variable,
with flux changes of a factor 5 or more.
About 15$\%$ of the sources in Table 1 show
this behavior. 
However not all the blue X-ray sources are variable and it is not possible with these two observations
to determine a proper variability time scale. Since most of
these sources do not vary significantly within the
observation intervals of 1999 Oct. and 1999 Nov., 
we conclude that we observe variability timescales on half a day and longer. 
Variability on shorter timescales may be possible (see Garmire et al. 2000), but in most cases the statistics per bin do not allow
for a conclusive result. The two lower panels in figure 2 (source 78 and 65) show typical light curves
obtained from these sources. In some cases we observe slightly enhanced, flare-like activity over 10 to 20 ks.
Also a small fraction ($< 5\%$) of the less absorbed sources show variability. In these sources we observe
giant flares, in which the X-ray flux changes from the detection limit to about a factor 10 to 30 above
within 1 to 3 ks, with decay times of 10 to 20 ks. The second panel in figure 2 (source 80) demonstrates
such a behavior.

\section{The Resolved Orion Trapezium}

As a subset of the observed star field we focus in the following sections on the Orion Trapezium and its 
immediate environment. The positions in Table 1 for the five Trapezium stars agree well with the positions
found recently by Simon et al. (1999).
Figures 1c and 1d show a 30 arcsec blowup from figures 1a and 1b around
the Trapezium. Some of the detected sources are too faint to accumulate a broadband spectrum and
may not show well in the Figure. We mark some of the sources with circles in the field in which
the detected number of counts is the larger of the two observations. Figure 3 shows a contour plot of the merged fields 
around the Trapezium. 
The results of the X-ray analysis are summarized in table
2. For the spectral fits as well as flux determinations
we used one or, where applicable, two-temperature Raymond-Smith model spectra.

\subsection{Diffuse Emission}

We searched for diffuse emission within the Trapezium. This is actually very difficult because of the 
dominance of the point spread function (PSF, from CIAO 1.1 release) of
$\theta^1 Ori$ C. Because of its proximity to Earth we
do not anticipate a contribution
from a scattering halo due to interstellar material.
From the measured column density in section 4.2
we predict an 0.8$\%$ contribution to the surface brightness (Predehl $\&$ Schmitt 1995). 
In order to estimate the fraction of a possible diffuse emission component within the field shown in figure 1c,
we removed all detected sources
within a radius of 2" of the detected PSF centroid, except for $\theta^1
Ori$ C, A and E, which are the dominating PSFs, 
and summed all counts. The count fraction
of the sum of $\theta^1 Ori$ A and E to $\theta^1 Ori$ C
is 50.4$\%$. We then removed the counts within a 2" radius in $\theta^1 Ori$ C, A and E as well,
and compared the remaining fraction to what one would expect from the modelled fractional encircled
energy function of a PSF for a point source. A comparison of
the measured to the modeled count fraction revealed equivalence within 1.5$\%$,
which is within the statistical uncertainties involved.    
Thus with high confidence we do not detect any diffuse
emission above our sensitivity limit.
    
\subsection{The Early Type Stars $\theta^1 Ori$ A-E: Emission from Stellar Winds}

The measured ACIS/HETGS zeroth-order
count rates range from 0.21 cts/s for the brightest component $\theta^1 Ori$ C, to 0.002 cts/s
for its faintest member  $\theta^1 Ori$ D. The three brighest members, C, E, and A, appear persistently bright
with X-ray luminosities (from table 1) between 1.85$\times 10^{32}$ \ergsec and 2.08$\times 10^{31}$ \ergsec.
Their light curves (figure 2 top, for $\theta^1 Ori$ C) do not show variability.
The ACIS spectra required more than one spectral component to give good fits. Here we applied a two-temperature 
Raymond-Smith model (fixed at solar abundances). The resulting
fits are remarkably similar for the three stars,
with all three showing a soft
component between 0.80 and 0.89 keV and a hard component between 2.18 and 3.15 keV. 
The flux ratios of hard flux over soft flux in E and A are
0.26 and 0.40, respectively, while in
C the same ratio is only 0.07. As a consequence 
of about 20$\%$ photon pileup in the CCD spectra, the results for $\theta^1 Ori$ C are somewhat uncertain. 
Here we refer to the detailed analysis of high resolution grating spectra from $\theta^1 Ori$ C (Schulz et al. 2000).   

An important result of our analysis is the
determination of the column density along the
line of sight towards
the Orion Trapezium. Here we make the
assumption that $\theta^1 Ori$ A, C, and E are foreground
objects projected onto L1641 and hence are not
significantly intrinsically absorbed, such 
that they are good representatives of the column density towards the Trapezium stars.
The measured column densities are very similar in all the sources and the average yields 1.93$\pm0.29 \times10^{21}$ cm$^{-2}$.
Given the foregoing assumption, we consider that sources that show
significantly higher absorption 
than this value are either intrinsically
absorbed or lie behind the Trapezium, such that they 
are embedded within L1641.

The X-ray emission from $\theta^1 Ori$ B and D have quite
different signatures compared with the other three stars. Their X-ray
fluxes and luminosities are an order of magnitude lower and they 
appear persistent within each observation, with luminosities between 1.94 and 1.51$\times 10^{30}$ \ergsec for 
$\theta^1 Ori$ B and 3.12 to 3.22$\times 10^{29}$ \ergsec for $\theta^1 Ori$ D.

The spectral fit results of $\theta^1 Ori$ B and $\theta^1 Ori$ D are very different as well.
$\theta^1 Ori$ B  
is probably the most peculiar system within the Trapezium
since it is a quadruple system (Petr et al. 1998). With a separation of 0.97" from optical observations, we clearly
see $\theta^1 Ori$ B as a double star (see figure 3), however with the dominating X-ray emission positioned on
component B1. Some emission extends onto component B, however since the flux fraction of the extended emission 
is less than 5$\%$ we can only provide the peak position as a detected source in table 1.
The spectral fit clearly rejects two components and
the column density could not be constrained during the fit. We then fixed the absorbing column 
at 1.38 $\times 10^{21}$ $cm^{-2}$, 
where the fit shows a very shallow minimium. To fix
this parameter near the value measured for the other Trapezium stars is justified, since $\theta^1 Ori$ B 
is an identified member of the Trapezium. Independent of the column density we find that
the spectral fits do not require a soft component $\theta^1 Ori$ A, C, and E.

$\theta^1 Ori$ D has the lowest observed flux of the five Trapezium stars. The fitted column density 
is slightly lower than observed in $\theta^1 Ori$ A, C, and E.
The spectral fit shows that, in contrast to the other members, no hard component is present.
The fit yields a temperature of kT = 0.7 keV. 
If we compare the flux of this spectrum to the soft flux fraction observed in $\theta^1 Ori$ A, C, and E,
we find that it is an order of magnitude lower.

\subsection{Young Stellar Objects and Proplyds}

From the analysis of the of the two observations it is clear that the core of the Orion Trapezium Cluster harbours a
large variety of hard, embedded, and variable sources. 
We detect several faint X-ray sources
projected within the Trapezium, as well as in its close vicinity. 

The X-ray light curve of source $\#$80 (figure 2) 
clearly shows a
quite powerful X-ray flare in 1999 October, with a rise time
of 2 ks and an exponential decay time of 20 ks. During the outburst the source shows an average X-ray luminosity of 2.2 $\times 10^{31}$ \ergsec,
with a peak luminosity of 5.5$\times 10^{31}$ \ergsec. Before the flare as well as throughout the second observation its
X-ray luminosity is about a factor 20 lower. We can only accumulate a spectrum during the flare. It shows 
a very high temperature of 4 keV. This outburst is quite reminiscent of
recent observations of hard flares in the Orion region (Yamauchi $\&$ Kamimura 1999) as well as the $\rho$ Oph Dark Cloud 
(Kamata et al. 1997) with \asca. 
The source is identified within 0.1" with the radio source GMR 25 (Garay et al. 1987) and was also seen at 2.2$\mu m$
by Simon et al. (1999). 
From the \asca observations it remained still uncertain whether class 0 or class I protostars
could be the origin of these flares. From the radio variability, Felli et al. 1993 classified the radio emission of this source
as non-thermal, which excludes a class 0 object, which should emit thermal radio emission only
(Carkner et al. 1998). Since no thermal radio emission has
been reported so far, and 
the persistent X-ray luminosity of this source (1.1$\times 10^{30}$ \ergsec) is quite strong, we conclude that this flare 
is associated with a Class II or III T Tauri star. 
      
GMR 3 is another radio source detected near the Trapezium, which coincides within 0.1" with source $\#$78. 
The light curve in figure 2 shows that it is persistently bright, showing some enhanced X-ray activity
at the end of the second observation.
This source appears deeply embedded with a column density of up to 2.61$\times 10^{22}$ cm$^{-2}$,
which is a factor 10 higher than expected for the
Trapezium. Its X-ray flux is relatively constant within the first
observation, but shows some variability in the second observation.
The spectrum can be fit by a similar single temperature
spectrum in both observations with temperatures of  kT =
1.65 and 1.85 keV, respectively.
It shows constant but weak thermal radio emission, which indicates a young
object. The star is visible in IR as
well as optical and
shows strong H$\alpha$ emission.
O'Dell $\&$ Wen (1994) identified this
object as a proplyd. 
In a survey of stellar properties of the ONC Hillenbrand derived an age for this star of 1.4$\times 10^5$ yr.
Given the properties above we conclude that the absorbed X-ray emission is associated with
a class II T Tauri star surrounded by a circumstellar disk.
 
Source $\#$64 is detected at 2.1$\mu m$ (McCaughrean $\&$ Stauffer 1994). It is also listed in the SIMBAD catalog
as [OW94]161-323, although O'Dell $\&$ Wen (1994) do not explicitly identify this position with a proplyd.
In X-rays it is highly variable, with an
X-ray luminosity of 7.7$\times 10^{30}$ \ergsec in the first, but below the detection threshold of 1.1
$\times 10^{29}$ \ergsec in the second observation. The source is clearly absorbed and hard, however we cannot 
properly determine a temperature.  
Source $\#$65 has very similar X-ray properties.  
It is identified with the thermal radio
source GMR 7. Bally et al. (1998)
describe this object as an externally illuminated young star-disk system ( [OW94]163-317 ). 
The light curve of source $\#$65 is shown in figure 2 at the bottom.
It does not
show much short term varibility, but it appears that the source changes in flux 
by more than an order of magnitude on timescales of at least one half day.

Sources $\#$58, 70, 72, and 74 were too faint to accumulate significant spectra though they appeared
well above the detection threshold at luminosities between
5$\times 10^{29}$ and 3$\times 10^{30}$ \ergsec, assuming an
absorbed  spectrum similar to that determined
for source $\#$ 78. Source $\#$58 is only identified with an IR position
(McCaughrean $\&$ Stauffer 1994), the others  with thermal radio sources 
GMR 21, 6, and 17 (Garay et al. 1987), respectively, as well as optical (O'Dell $\&$ Wen 1994), and IR (McCaughrean $\&$ Stauffer 1994)
sources. Except for $\#$58, they appear variable in our two observations, i.e are only detected in one of them, and
O'Dell $\&$ Wen (1994) identify proplyds within 0.3" of these positions.  

In order to emphasize the strong correlation of the X-ray emission in the Trapezium with the optical emission
observed with the Hubble Planetary Camera in figure 3 we
overlay our X-ray data onto data from Figure 2 in Bally et al. (1998),
which is a composite image from narrow band filter
observations of N II, H$\alpha$, and O I. 
$\theta^1 Ori$ B shows emission from both components, with B1 clearly dominating.  
The overlaid X-ray contours clearly show that many of
the proplyds are X-ray
sources (e.g., sources 
65, 70, 72, 74);
although source 64 clearly cannot be
associated with 161-322. The overlay also shows that many of
the other proplyds in the vicinity do not show X-rays;
however,
if we lower our threshold to 3$\sigma$, we find enhanced X-ray emission 
coincident with at least two more proplyds. 

\section{Summary and Discussion}

Our two observations of the Orion Trapezium Cluster in 0th order of the Chandra HETGS with a spatial resolution
of 0.5" thoughout almost the entire field of view reveal many new details about the X-ray properties of the
Cluster. They allow us to resolve the Trapezium Cluster entirely into point sources, study their variability patterns
and perform detailed X-ray spectroscopy.  
Here we report on the following results:

We resolved all the emission around the Trapezium into point sources, detecting 111 X-ray sources.
The cluster
shows no diffuse emission above the total (both observations merged) sensitivity limit of 2.0$\times10^{28}$ \ergsec
per arcsec$^2$. Besides the five main
Trapezium stars we detected six additional X-ray point sources in the close vicinity of the Trapezium. 
Most sources are faint. Only 12 sources 
have persistent X-ray luminosities above $10^{31}$
\ergsec; four of these are part of the Trapezium and are
identified with early type stars. The avarage luminosity of the fainter X-ray sources is
around 3$\times 10^{30}$ \ergsec, in agreement with the result from Garmire et al. (2000) for solar mass
PMS stars.  

The two observations showed considerable variability on timescales larger than 0.5 days. 
Fifteen percent of the sources vary by more than a factor 5 in luminosity; the largest excursion 
in luminosity was from 1.3$\times 10^{31}$ \ergsec to 4.1$\times 10^{29}$ \ergsec .   
Most of these variable sources appear absorbed with column densities between 1.7$\times 10^{21}$ cm$^{-2}$ to 
1.1$\times 10^{24}$ cm$^{-2}$ with standard deviations of 5$\%$ and 10$\%$ respectively.
The three bright Trapezium stars display very similar values
of line of sight column density,  
with a mean value of $N_H =$ 1.93$\pm0.29 \times 10^{21} cm^{-2}$.  

The most luminous source is $\theta^1$ Ori C with L$_x$ =  1.85$\pm0.18 \times 10^{32}$ \ergsec. This
is an order of magnitude lower than the estimated luminosity from \asca of $10^{33}$ \ergsec. If we integrate
the luminosities of all sources within the \asca point spread function, however, we get a reasonably consistent    
result. 
Stahl et al. (1993) measured a period of 15.4 days in optical and UV data, which was also seen
in the \ros HRI data (Caillault et al. 1994). From the ephemeris given by Stahl et al. we determine
phases of 0.68 for our first, and 0.23 for our second observation. From the HRI peak luminosity of
given in Caillault et al. we expect luminosites of 2.3$\times 10^{32}$ \ergsec and 2.0$\times 10^{32}$ \ergsec,
respectively, for our observations. Although the results in Table 1 are slightly lower
than these estimates, they are not inconsistent given the various systematic uncertainties.

\subsection{Early Type Stars}

Efforts to model X-ray spectra from stellar winds
have been quite modest since the first observations with \ein (Seward et al. 1979). The main
reason is that the theory of line radiation-driven stellar winds, which explains successfully
the mass loss from a hot luminous star from UV data, 
could never correctly predict observed X-ray fluxes. 
Therefore to date no definitive models for the production of X-rays in stellar winds exist. 
The X-ray emission is now thought to be produced by shocks forming from instabilities within a
radiatively driven wind (Lucy $\&$ White 1980), where hot blobs plow through rarified cool wind material
well into the terminal flow of the wind (Lucy 1982). This model produces X-rays of about 0.5 keV temperature.
Hellier et al. (1993)
fitted high signal-to-noise PSPC spectra of $\zeta$ Pup with detailed NLTE models under the assumption that
the X-rays arise from shocks distributed throughout the wind and allowing for recombination in the outer
regions of the stellar wind. The best fits predicted two temperatures of 1.6 and 5.0$\times10^6$ K with shock velocities
around 500 km s$^{-1}$. 

We performed detailed spectroscopy on the Trapezium stars as well as for most of the nearby X-ray
sources.  $\theta^1$ Ori A and E are the brightest sources besides C, with persistent X-ray luminosities above
$10^{31}$ \ergsec. None of these sources are observed to be variable by more
than 10$\%$. All three sources are identified with massive, very early type stars: 
$\theta^1 Ori$ C is a massive ($\sim 25$ \Msun) O7V star with a recently discovered 
low-mass companion (Weigelt et al. 1999);
$\theta^1 Ori$ A is identified with an eclipsing binary of 0.2" separation and a 63.43 day period;
$\theta^1 Ori$ E is a massive early B0.5 type star, and so far no companion has been detected. 
The main component of $\theta^1 Ori$ A
is not very well determined and speculations range from spectral type B0V to B1V (Lloyd  $\&$ Strickland 1999).
The nature of the companion is similarly unclear and
identifications range from a low-mass pre-main sequence star (Strickland $\&$ Lloyd 2000) to a B8-AOV type star
(Vitrichenko 1999). There are indications that $\theta^1 Ori$ A is even a triple system (Petr et al. 1998).
The spectra of $\theta^1 Ori$ A, C, and E were fitted with a two-temperature 
Raymond-Smith spectrum of kT $\sim$ 0.8 keV and kT $\sim$ 2-3 keV.
Since these sources are identified with very early type, massive stars we conclude that the X-ray spectra 
show X-ray emission characteristics of strong stellar winds. Specifically, in the case of
$\theta^1 Ori$ A it is more likely that the X-rays originate from its very early type main component
and not from any companion. In fact if the companion is a T Tauri star, its X-ray spectrum may only contribute
a small fraction of the observed flux. 
If it is a late type B-star, we would not expect any X-rays at all (Bergh\"ofer $\&$ Schmitt 1994).
The spectral fits also agree with the results from previous \asca observations (Corcoran et al. 1994, Cohen et al. 1996)
that X-rays from early type stars have two or more temperature components.

The sources $\theta^1$ Ori B and D have luminosities well below $10^{31}$ \ergsec.
The fainter one, $\theta^1$ Ori D, shows only a soft component of 0.7 keV and we may speculate
that here the wind lacks the hot component observed in $\theta^1$ Ori A, C, and  E, which are of earlier type.
The \ros spectrum of $\lambda$Sco and $\alpha$ Vir, which are of similar spectral type (B1V), show similar luminosities
and spectral temperatures. Here we may observe an effect, that with increasing spectral type not only the 
X-ray luminosity decreases but also the spectrum softens.
$\theta^1$ Ori B is brighter, lacks the soft component, and shows
a hot component. The second member of $\theta^1$ Ori B, $\theta^1$ Ori B1, whose stellar type is not 
precisely known,  seems to be the dominating source for the observed X-rays. 
Star B is identified 
with a spectral type of B3V (Abt et al. 1991), which may already be late enough to have such a weak wind that it cannot produce
the X-ray luminosities observed. \ros observations indicate that in B-stars of later type, the 
luminosity decreases. Cassinelli et al. (1994) analysed  several
B-stars of this particular spectral type and found soft X-ray luminosities 
of the order of $10^{28}$ \ergsec and lower, which is well below our sensitivity
limit. On the other hand, star B1 is thought to be of earlier type (B1) and \ros observations predict X-ray luminosities of
above $10^{30}$ \ergsec, which we do not observe. It also remains unclear why we observe such hard spectrum.

The spectra of the early type stars in the Trapezium seem unusually hot compared to what stellar wind models
currently predict. Four of the five main components, A, B, C, and E, show spectral temperatures between
2.4$\times10^7$ K and 3.7$\times10^7$ K, which is an order of magnitude hotter than current stellar wind models
predict. In the case of $\theta^1$ Ori C, Schulz et al. (2000) recently determined a temperature range of 0.5 to 6.1$\times10^7$ K
from X-ray emission lines. These temperatures were deduced from H-like ion 
species up to S XVI and Fe XXV, which  rules out a significant inverse Compton continuum as suggested by Chen $\&$
White (1991).
Babel $\&$ Montmerle (1997a) presented the hypothesis that in the 
case of $\theta^1$ Ori C a dipolar magnetic field is embedded in the radiation-driven wind and X-rays are 
produced by a magnetically confined wind shock. In the case of the hot A0p star IQ Aur,  Babel $\&$ Montmerle (1997b)
modelled temperatures up to 9$\times10^6$ K by including the effects of magnetic confinement. If in addition substantial
stellar rotation was included, temperatures could be as high as 
3 $\times10^7$ K. Therefore it seems compelling to include magnetic field effects as well as centrifugal
forces into stellar wind models for these hot stars. The fact that such high temperatures are present
in $\theta^1$ Ori A, B, and E, as well as in $\tau$Sco (Cohen et al. 1997), also shows that these effects may be 
more common in early type stars.   

\subsection{YSOs}
  
Although it is now widely accepted that X-ray emission from young stars emerges from  
hot stellar coronae heated via a stellar dynamo, the underlying
physics of this mechanism is still poorly understood. High levels
of magnetic activity in connection with rotation, accretion, and/or outflows may play a key role 
(see Feigelson $\&$ Montmerle 1999 for 
a review). Recent imaging observations showed that the X-rays of most T Tauri stars vary on time scales of
days and a substantial number of stars display high amplitude flares with time scales of hours. X-ray spectra
show  more than one temperature and are usually modelled with soft spectral components of 2-5$\times10^6$ K and some
hard components during flares of up to 3 $\times10^7$ K. Montmerle et al. (2000) recently modelled the X-ray emission
from a giant flare with a peak temperature of 7 $\times10^7$ K in YLW 15 observed with \asca (Tsuboi et al. 2000)   
with strong magnetic shearing and reconnection between the central star and the accretion disk. In the case
of WL 6 (Kamata et al. (1997) the X-ray flaring is rotationally modulated (see also Stelzer et al. 1999). 
Montmerle et al. (2000) inferred a mass-rotation relation, in which higher mass stars are fast rotators and
giant hard flares are generated through strong star-disk magnetic interactions, while lower mass stars are slower and
and their X-ray emission is solely of stellar origin.  
    
The vicinity of the Orion Trapezium harbours a large variety of YSOs. Most notable are the large number of proplyds
detected with the Hubble Space Telescope (O'Dell et al. 1993). 
We detected 8 faint X-ray sources, where six sources are identified
with radio objects, five sources with proplyds in the vicinity of $\theta^1$ Ori C 
(i.e. proplyds 163-317, 166-316, 167-317, 168-328, and 171-334). 
One source is identified with a YSO in the the IR only. These X-ray sources
show luminosity variations between a factor 3 and 10, and
have spectra indicating that they are heavily absorbed.
The X-ray emission, together with optical, IR and radio surveys (Bally et al. 1998, McCaughrean and Stauffer 1994,
Felli et al. 1993) indicate that these sources are embedded Class I, Class II,  or Class III protostars. We do not believe that
the X-rays are generated through induced shocks by the wind $\theta^1$ Ori C. First, the 
flux of these sources varies by large factors which we would not expect under constant illumination
unless the density structure in the wind shock at the proplyds changes similarly. That would imply similar
brightness variations in the optical, but no such variations have been reported so far. On the other hand, in source
$\#$78 we see variability timescales of the order of 10 min to 2 hours (figure 2). 
From this we deduce an upper limit of the  size of the emitting 
region of the order of 1.0 - 12 AU, which is too small compared
to size of the proplyds ($\sim$  500 AU, Bally et
al. 1998). Second, the 
spectra we analysed are intrinsically  absorbed. Thus the emission likely originates from deep inside
the dusty disks, rather than from the wind shock region.

In the context of recent observations of hard X-ray emission from YSOs, these results are quite remarkable; 
the light curves indicate that only one of the sources actually displays large flaring activity, and the
others are persistent in flux with some minor activity. Very large flux variations occur only on
time scales longer than 12 h. The spectra, however, indicate extremely high temperatures at quite 
modest luminosities, from which we conclude that the X-ray emission may not be related to star-disk
magnetic interactions in connection with rapid rotation as proposed by Montmerle et al. (2000). The
observed X-ray behavior of these proplyds seem more related to recent \asca observations of SU Aurigae
(Skinner $\&$ Walter 1998), which has been identified with a type G2  CTTS and where it was suggested 
that the X-ray emission arises from a magnetically confined stellar plasma.
In this respect long term X-ray observations of these proplyds will not only enhance the significance
of these hot spectra but also  allow a search for rotational modulations and lead to better understanding
of the magnetic properties and their role in the evolution of these protostars. 

\acknowledgements

The authors thank John Bally for generously providing us with the HST PC image of the Trapezium.
We also thank all the members of the \chandra team for their enormous efforts. This research is funded
in part by the Smithonian Astrophysical Observatory contract SV-61010 (CXC) and NAS8-39073 (HETG) 
under the Marshall Space Flight Center.

\newpage

\begin{table}
\footnotesize
\caption{SOURCE POSITIONS, COUNT RATES AND IDENTIFICATIONS.}
\tablenum{1}
\begin{tabular}{lccccc}
\tableline
\tableline
$\#$ & $\alpha, \delta (2000)$ & total cts  & Oct. 99 & Nov. 99 & ID       \\
     & $5^h35^m$,$-5^{\circ}$  & (Oct/Nov)  &$10^{-3}$~cts/s &$10^{-3}$~cts/s \\
\tableline
A & $15^s.82, 23'14".19$ & 3014 & 35.65 & 39.02 & $\theta^{1}$ Ori A \\
B & $16^s.07, 23'06".99$ & 263 & 3.34 & 3.03  & $\theta^{1}$ Ori B \\
C & $16^s.46, 23'22".89$ & 16640 & 205.64 & 201.31  & $\theta^{1}$ Ori C \\
D & $17^s.25, 23'16".53$ & 152 & 1.77 & 2.01  & $\theta^{1}$ Ori D \\
E & $15^s.77, 23'09".86$ & 5372 & 61.50 & 72.84  & $\theta^{1}$ Ori E \\
\tableline
6 & $11^s.73, 23'40".50$ & 37 & 0.34 & 0.64  & [H97b] 9008,[PSH94] 8 \\
7 & $12^s.29, 23'48".06$ & 490 & 4.16 & 8.97  & Parenago 1825 \\
8 & $12^s.35, 22'41".34$ & 28 & 0.10 & 0.73  & [AD95] 2736 \\
9 & $12^s.41, 23'51".36$ & 41 & 0.66 & 0.26  & ... \\
10 & $12^s.58, 23'01".96$ & 47 & 0.52 & 0.67  & [PSH94] 18 \\
11 & $12^s.61, 23'44".19$ & 51 & 0.60 & 0.67  & Parenago 1807 \\
12 & $12^s.97, 23'54".76$ & 72 & 0.62 & 1.31  & [AD95] 3132 \\
13 & $12^s.99, 23'30".28$ & 14 & 0.14 & 0.22  & [PSH94] 44,[JW88] 412 \\
14 & $13^s.22, 22'55".03$ & 44 & 0.30 & 0.93  & [AD95] 2717 \\
15 & $13^s.31, 22'39".13$ & 14 & 0.16 & 0.19  & [GWV98] 053246.204-052432.86 \\
16 & $13^s.45, 23'40".29$ & 171 & 1.01 & 3.83  & [AD95] 3141 \\
17 & $13^s.54, 23'30".98$ & 62 & 0.72 & 0.83  & [AD95] 3255 \\
18 & $13^s.59, 23'55".40$ & 81 & 0.52 & 1.76  & [AD95] 3132 \\
19 & $14^s.06, 23'38".48$ & 37 & 0.36 & 0.61  & Parenago 1823 \\
20 & $14^s.08, 22'22".10$ & 65 & 0.94 & 0.57  & [M77] HH08, [S99] 24 \\
21 & $14^s.08, 22'36".43$ & 67 & 0.52 & 1.31  & [GWV98] 053246.628-052429.82 \\
22 & $14^s.29, 23'04".17$ & 61 & 0.68 & 0.86  & [MS94] 3, [S99] 27 \\
23 & $14^s.27, 24'24".73$ & 26 & 0.52 & - & Parenago 1826,V$\*$ V1328 Ori \\
24 & $14^s.32, 23'08".31$ & 342 & 2.01 & 7.70  & [H97b] 9061, [S99] 28 \\
25 & $14^s.36, 22'32".66$ & 120 & 1.49 & 1.44  & [S99] 32 \\
26 & $14^s.36, 22'54".02$ & 38 & 0.34 & 0.67  & Parenago 1842 \\
27 & $14^s.37, 23'33".13$ & 122 & 1.13 & 2.08  & Parenago 1843 \\
28 & $14^s.40, 22'36".48$ & 29 & 0.28 & 0.48  & [JGS92] 2 \\
29 & $14^s.50, 22'38".66$ & 103 & 1.41 & 1.02  & [RLK73] IRc4 (?) \\
30 & $14^s.55, 23'16".01$ & 313 & 0.76 & 8.78  & [MS94] 13, [S99] 49 \\
31 & $14^s.57, 24'07".93$ & 22 & 0.34 & 0.16  & ... \\
32 & $14^s.65, 22'33".67$ & 310 & 4.32 & 2.97  & [S99] 38 \\
33 & $14^s.67, 23'01".75$ & 253 & 4.64 & 0.64  & [S99] 37 \\
34 & $14^s.69, 22'49".60$ & 25 & 0.28 & 0.35  & [MS94] 11, [S99] 41 \\
35 & $14^s.72, 23'22".83$ & 59 & 0.60 & 0.93  & [S99] 69 \\
36 & $14^s.73, 22'29".70$ & 809 & 14.78 & 2.11  & [S99] 46 \\
37 & $14^s.80, 24'07".13$ & 15 & 0.22 & 0.13  & [AD95] 2626 \\
38 & $14^s.83, 23'46".47$ & 40 & 0.44 & 0.57  & Parenago 1845,[MS94] 21,[JW88] 464 \\
39 & $14^s.87, 22'31".44$ & 27 & 0.40 & 0.22  & [S99] 53 \\
40 & $14^s.91, 22'39".14$ & 1090 & 17.33 & 6.99  & [S99] 56 \\
41 & $14^s.89, 22'25".39$ & 140 & 1.51 & 2.04  & [S99] 78 \\
42 & $14^s.92, 24'12".83$ & 42 & 0.48 & 0.57  & [H97b] 9080,[PSH94] 80 \\
43 & $14^s.95, 23'39".26$ & 170 & 2.13 & 2.01  & Parenago 1844 \\
44 & $15^s.09, 22'31".52$ & 39 & 0.54 & 0.38  & [H97b] 9086,[PSH94] 86, [S99] 53 \\
45 & $15^s.16, 22'17".40$ & 138 & - & 4.41  & [GWV98] 053247.641-052410.31 \\
46 & $15^s.19, 22'24".04$ & 96 & 1.01 & 1.44  & [S99] 67, [HAB84] 63 \\
47 & $15^s.19, 22'54".12$ & 69 & 0.92 & 0.73  & Parenago 1841 \\
48 & $15^s.26, 22'56".83$ & 897 & 10.60 & 11.62  & Parenago 1862,V$\*$ V348 Ori \\
49 & $15^s.34, 22'25".32$ & 37 & 0.30 & 0.70  & [H97b] 9096, [PSH94] 96 \\
50 & $15^s.34, 22'15".47$ & 365 & - & 11.66  & Parenago 1838\\
51 & $15^s.38, 23'33".57$ & 56 & 0.66 & 0.73  & [MS94] 29\\
52 & $15^s.44, 23'45".46$ & 50 & 0.76 & 0.38  & [S99] 77, [OW94] 154-346        \\
53 & $15^s.49, 22'48".53$ & 57 & 0.82 & 0.51  & [S99] 81\\
54 & $15^s.60, 24'03".06$ & 54 & 0.70 & 0.61  & [AD95] 3125 \\
55 & $15^s.63, 22'56".44$ & 1996 & 26.26 & 21.59  & Parenago 1862,V$\*$ V348 Ori \\
56 & $15^s.68, 23'39".15$ & 108 & 0.80 & 2.17  & [MS94] 37, [S99] 87 \\
\tableline
\end{tabular}
\normalsize
\end{table}

\newpage

\begin{table}
\footnotesize
\caption{SOURCE POSITIONS, COUNT RATES AND IDENTIFICATIONS.}
\tablenum{1}
\begin{tabular}{lccccc}
\tableline
\tableline
$\#$ & $\alpha, \delta (2000)$ & total cts  & Oct. 99 & Nov. 99 & ID       \\
     & $5^h35^m$,$-5^{\circ}$  & (Oct/Nov)  &$10^{-3}$~cts/s &$10^{-3}$~cts/s \\
\tableline
57 & $15^s.76, 23'38".40$ & 44 & 0.12 & 1.21  & [MS94] 41, [S99] 90\\
58 & $15^s.82, 23'11".94$ & 52 & 0.62 & 0.67  & [S99] 93 \\
59 & $15^s.84, 22'45".92$ & 39 & 0.36 & 0.67  & Parenago 1860 \\
60 & $15^s.97, 23'49".70$ & 751 & 6.09 & 14.21  & [S99] 104 \\
61 & $16^s.07, 23'53".45$ & 73 & 0.76 & 1.12  & Parenago 1870,V$\*$ AC Ori \\
62 & $16^s.08, 24'11".61$ & 20 & 0.30 & 0.16  & [AD95] 3123 \\
63 & $16^s.07, 22'54".16$ & 44 & 0.58 & 0.48  & [MS94] 55, [S99] 114 \\
64 & $16^s.10, 23'23".11$ & 310 & 5.07 & 1.76  & [OW94] 161-323, [S99] 113 \\
65 & $16^s.29, 23'16".50$ & 215 & 3.22 & 1.69  & [BSD98] 6,[JW88] 512,[MS94] 63 \\
66 & $16^s.37, 24'03".33$ & 337 & 4.28 & 3.90  & ... \\
67 & $16^s.49, 22'56".30$ & 21 & 0.14 & 0.45  & [MS94] 69, [S99] 137 \\
68 & $16^s.49, 22'35".24$ & 21 & 0.16 & 0.42  & [S99] 138, H97b] 519a \\
69 & $16^s.58, 24'06".11$ & 119 & 0.88 & 2.40  & V $\*$V1279 Ori, Parenago 1869 \\
70 & $16^s.61, 23'16".04$ & 104 & 1.19 & 1.41  & [BSD98] 7,[OW94] 166-316,[MS94] 70 \\
71 & $16^s.74, 22'31".10$ & 52 & 0.42 & 0.99  & [H97b] 9151,[OW94] 167-231, [S99] 145 \\
72 & $16^s.74, 23'16".36$ & 100 & 1.27 & 1.15  & $\theta^1$ Ori G,Parenago 1890 \\
73 & $16^s.76, 24'04".32$ & 604 & 7.84 & 6.71  & [H97b] 526b,[PSH94] 150 \\
74 & $16^s.77, 23'28".03$ & 45 & 0.40 & 0.80  & [BSD98] 5,[MS94] 74, [S99] 144\\
75 & $16^s.88, 22'22".35$ & 113 & 1.41 & 1.34  & [S99] 153 \\
76 & $16^s.00, 22'32".95$ & 413 & 5.15 & 4.92  & [S99] 103 \\
77 & $17^s.06, 23'39".74$ & 66 & 0.76 & 0.89  & [S99] 160, [AD95] 3144\\
78 & $17^s.06, 23'34".09$ & 779 & 7.80 & 12.36  & Parenago 1893 \\
79 & $17^s.38, 24'00".24$ & 48 & 0.50 & 0.73  & [AD95] 3130 \\
80 & $17^s.46, 23'21".08$ & 881 & 17.01 & 0.83  & [H97b] 9180,[MS94] 93, [S99] 175 \\
81 & $17^s.55, 22'56".73$ & 58 & 0.38 & 1.25  & [H97b] 553a,[PSH94] 185 \\
82 & $17^s.77, 23'42".63$ & 89 & 0.86 & 1.47  & [S99] 192, [H97b] 9195,[PSH94] 195,[MS94] 99 \\
83 & $17^s.81, 23'15".70$ & 18 & 0.10 & 0.42  & [OW94] 178-316S, [S99] 195 \\
84 & $17^s.87, 23'02".99$ & 72 & 0.44 & 1.60  & [H97b] 9201,[PSH94] 201,[MS94] 103 \\
85 & $17^s.94, 22'45".42$ & 4222 & 29.78 & 87.02  & [AD95] 3168 \\
86 & $17^s.96, 23'35".50$ & 40 & 0.60 & 0.32  & [MS94] 105 \\
87 & $18^s.03, 24'03".21$ & 65 & 0.92 & 0.61  & [H97b] 9210,[PSH94] 210 \\
88 & $18^s.05, 24'01".07$ & 56 & 0.68 & 0.70  & ... \\
89 & $18^s.19, 23'35".95$ & 165 & 2.57 & 1.15  & [AD95] 3271, [OW94] 182-336\\
90 & $18^s.32, 24'04".95$ & 39 & 0.50 & 0.45  & V$\*$ V1336 Ori \\
91 & $18^s.36, 22'37".38$ & 1135 & 13.73 & 14.21  & V$\*$ V1229 Ori,Parenago 1925 \\
92 & $18^s.66, 23'13".90$ & 150 & 2.17 & 1.31  & [HHM94]12 \\
93 & $18^s.69, 22'56".82$ & 155 & 1.69 & 2.24  & [H97b] 598a,[PSH94] 235 \\
94 & $18^s.95, 22'18".52$ & 40 & - & 1.28  & [S99] 204 \\
95 & $19^s.06, 23'49".75$ & 30 & 0.40 & 0.32  & V$\*$ V1337 Ori \\
96 & $19^s.11, 23'27".02$ & 25 & 0.14 & 0.57  & [S99] 235 \\
97 & $19^s.20, 22'50".63$ & 828 & 1.61 & 23.85  & [H97b] 9250,[MS94] 123,[PSH94] 250 \\
98 & $19^s.34, 23'40".56$ & 14 & 0.20 & 0.13  & ... \\
99 & $19^s.59, 23'57".32$ & 74 & 0.72 & 1.21  & [AD95] 2647 \\
100 & $19^s.79, 22'21".55$ & 233 & 4.58 & 0.10  & ... \\
101 & $20^s.17, 23'08".59$ & 111 & 0.40 & 2.91  & [H97b] 3075, [AD95] 3270 \\
102 & $20^s.45, 23'29".84$ & 152 & 1.55 & 2.36  & [H97b] 648b, [PSH94] 62, [AD95] 3151 \\
103 & $20^s.53, 24'20".86$ & 18 & 0.36 & - & [AD95] 3306 \\
104 & $20^s.66, 23'53".09$ & 60 & 0.72 & 0.77  & [AD95] 3135 \\
105 & $20^s.89, 22'34".32$ & 19 & 0.18 & 0.32  & [S99] 247 \\
106 & $20^s.90, 23'21".83$ & 191 & 0.46 & 5.36  & [H97b] 9271 \\
107 & $21^s.01, 23'55".59$ & 24 & 0.08 & 0.64  & [H97b] 9272, [PSH94] 72 \\
108 & $21^s.03, 23'48".00$ & 1947 & 17.43 & 34.20  & [H97b] 9287, [PSH94] 78 \\
109 & $21^s.22, 22'59".42$ & 27 & 0.22 & 0.51  & [AD95] 2720, [H97b] 9280, [PSH94] 280 \\
110 & $21^s.31, 24'11".47$ & 28 & 0.30 & 0.42  & V$\*$ V1231 Ori \\
111 & $21^s.35, 23'45".36$ & 314 & 5.65 & 0.96  & [PSH94] 279 \\
\tableline
\end{tabular}
\vskip 0.3cm
[AD95] $=$ Ali \&
Depoy 1995; [H97] $=$ Hildebrand 1997; [JW88] $=$  Jorgensen \&
Westerlund 1988; [GWV98] $=$  Gaume et al.\ 1998;
[M77] $=$  Munch 1977; [RLK73] $=$ Rieke, Low, \& Kleinmann 1973;
[HAB84]  $=$ Hyland et al.\ 1984; [OW94] $=$ O'Dell \& Wen
1994; [BSD98] $=$ Bally et
al.\ 1998; [HHM94] $=$ Hayward, Houck, \& Miles 1994; [MS94]
$=$ McCaughrean \& Stauffer 1994; [PSH94] $=$  Prosser et
al.\ 1994; [S99] $=$ Simon, Close, \& Beck 1999
\normalsize
\end{table}

\newpage

\begin{table}
\footnotesize
\caption{SPECTRAL PARAMETERS FOR IDENTIFIED TRAPEZIUM SOURCES.}
\tablenum{2}
\begin{tabular}{lccccccc}
\tableline
\tableline
     src$\#$ & N$_H ^{(1)}$& kT$_1$ & kT$_2$ & f$_x ^{(2)}$ & L$_x ^{(3)}$ & dof & $\chi^2$ \\
         &  & keV & keV & & & &   \\
\tableline
\tableline
      A  & 2.19$\pm$0.04 & 0.87$\pm$0.02 & 2.18$\pm$0.12 & 1.25$\pm$0.03 &  2.08$\pm$0.07 & 314 & 0.78 \\
         & 1.49$\pm$0.02 & 0.85$\pm$0.04 & 2.37$\pm$0.17 & 1.26$\pm$0.05 &  2.35$\pm$0.11 & 314 & 0.62 \\
      B  & 1.38          &               & 3.16$\pm$0.58 & 0.12$\pm$0.01 &  0.19$\pm$0.02 &  57 & 0.48 \\
         & 1.38          &               & 2.04$\pm$0.42 & 0.09$\pm$0.01 &  0.15$\pm$0.02 &  57 & 0.36 \\
      C$^{(4)}$  & 1.83$\pm$0.16 & 0.83$\pm$0.02 & 3.15$\pm$0.27 &12.50$\pm$0.13 & 18.52$\pm$1.80 & 312 & 1.32 \\
         & 2.03$\pm$0.24 & 0.80$\pm$0.03 & 3.20$\pm$0.37 &11.59$\pm$0.19 & 17.32$\pm$2.31 & 312 & 1.07 \\
      D  & 1.65          & 0.76$\pm$0.08 &               & 0.05$\pm$0.01 &  0.03$\pm$0.01 & 182 & 0.16 \\
         & 1.65$\pm$0.05 & 0.67$\pm$0.07 &               & 0.04$\pm$0.01 &  0.03$\pm$0.01 & 182 & 0.11 \\
      E  & 1.97$\pm$0.03 & 0.89$\pm$0.02 & 3.13$\pm$0.15 & 2.58$\pm$0.05 &  4.46$\pm$0.12 & 319 & 0.90 \\
         & 2.09$\pm$0.03 & 0.85$\pm$0.03 & 2.50$\pm$0.12 & 2.51$\pm$0.07 &  4.35$\pm$0.16 & 319 & 0.89 \\
     64  & 5.87$\pm$0.19 &               &32.3 $\pm$61.7 & 0.28$\pm$0.02 &  0.42$\pm$0.03 & 130 & 0.51 \\
         & 5.87          &               &               & 0.06$\pm$0.01 &  0.09$\pm$0.02 &     &      \\
     65  & 6.05$\pm$2.35 &               &31.8 $\pm$41.5 & 0.13$\pm$0.01 &  0.20$\pm$0.01 &  65 & 0.59 \\
         & 6.05          &               &               & 0.02$\pm$0.01 &  0.03$\pm$0.01 &     &      \\
     78  & 23.6$\pm$ 3.4 &               & 1.65$\pm$0.17 & 0.34$\pm$0.02 &  1.93$\pm$0.37 & 105 & 0.60 \\
         & 26.1$\pm$ 5.3 &               & 1.85$\pm$0.18 & 0.47$\pm$0.03 &  2.37$\pm$0.42 & 105 & 0.55 \\
     80  & 3.34$\pm$0.05 &               & 4.28$\pm$0.44 & 0.75$\pm$0.02 &  1.25$\pm$0.06 & 182 & 0.81 \\
         & 3.34          &               &               & 0.04$\pm$0.01 &  0.11$\pm$0.02 &     &      \\
\tableline
\end{tabular}
\break
$\rm ^1${in units of 10$^{21} cm^{-2}$}\\
$\rm ^2${in units of 10$^{-12}$ \ergcm}\\
$\rm ^3${in units of 10$^{31}$ \ergsec}\\
$\rm ^4${spectra affected by approx. 20$\%$ pileup}\\
\normalsize
\end{table}

\newpage

\begin{figure*}
\centerline{\epsfxsize=17.0cm\epsfbox{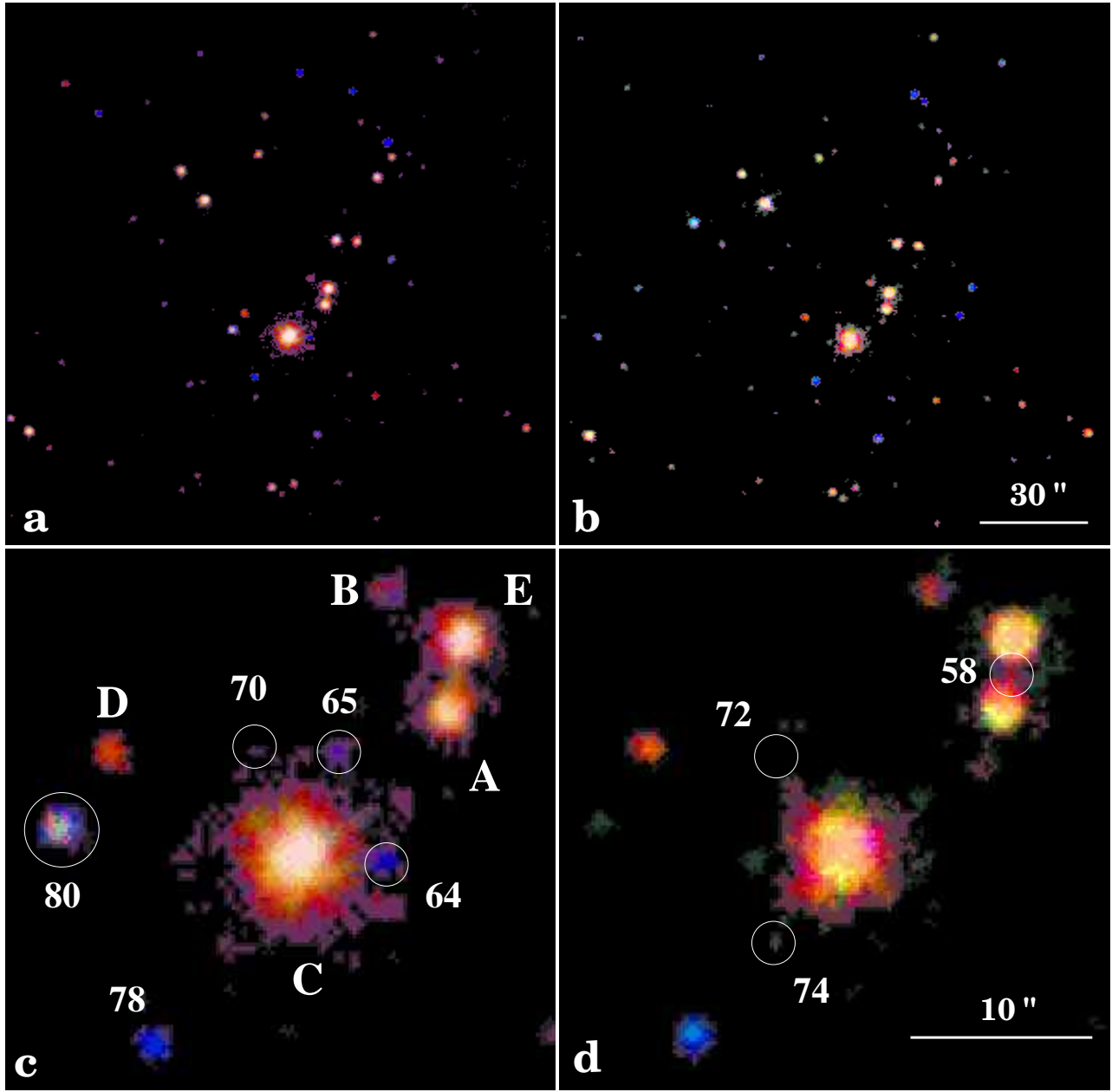}}
\figcaption{X-ray Variability in the Orion Trapezium. (a) and (b) show color coded images of a 160"x160" field
around the Trapezium, which are 3 weeks apart. (c) and (d) show the same for a close-up on the Trapezium.
The circles indicate the positions of detected sources in the frame where they are brightest.}
\end{figure*}

\begin{figure*}
\centerline{\epsfxsize=8.5cm\epsfbox{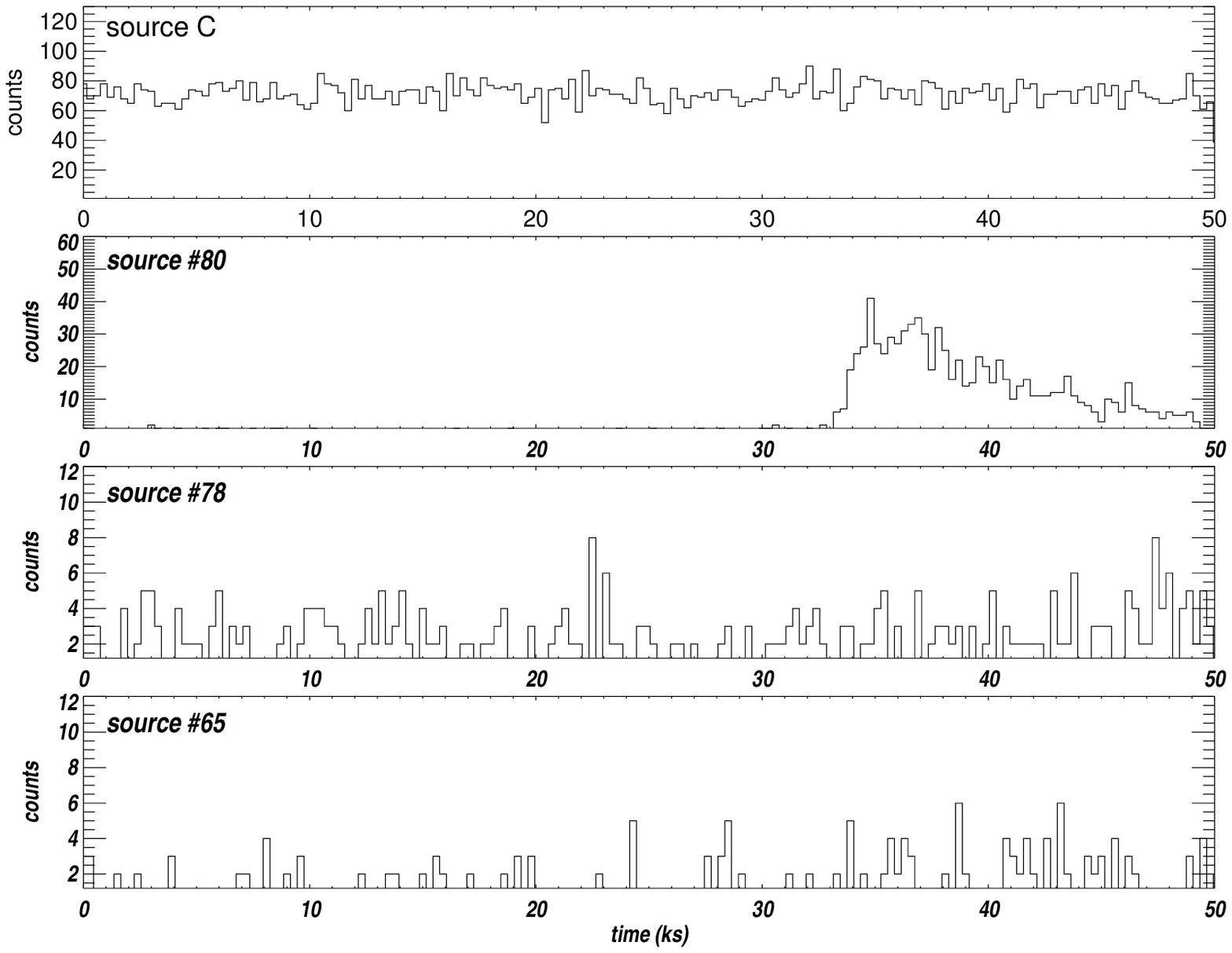}\epsfxsize=8.5cm\epsfbox{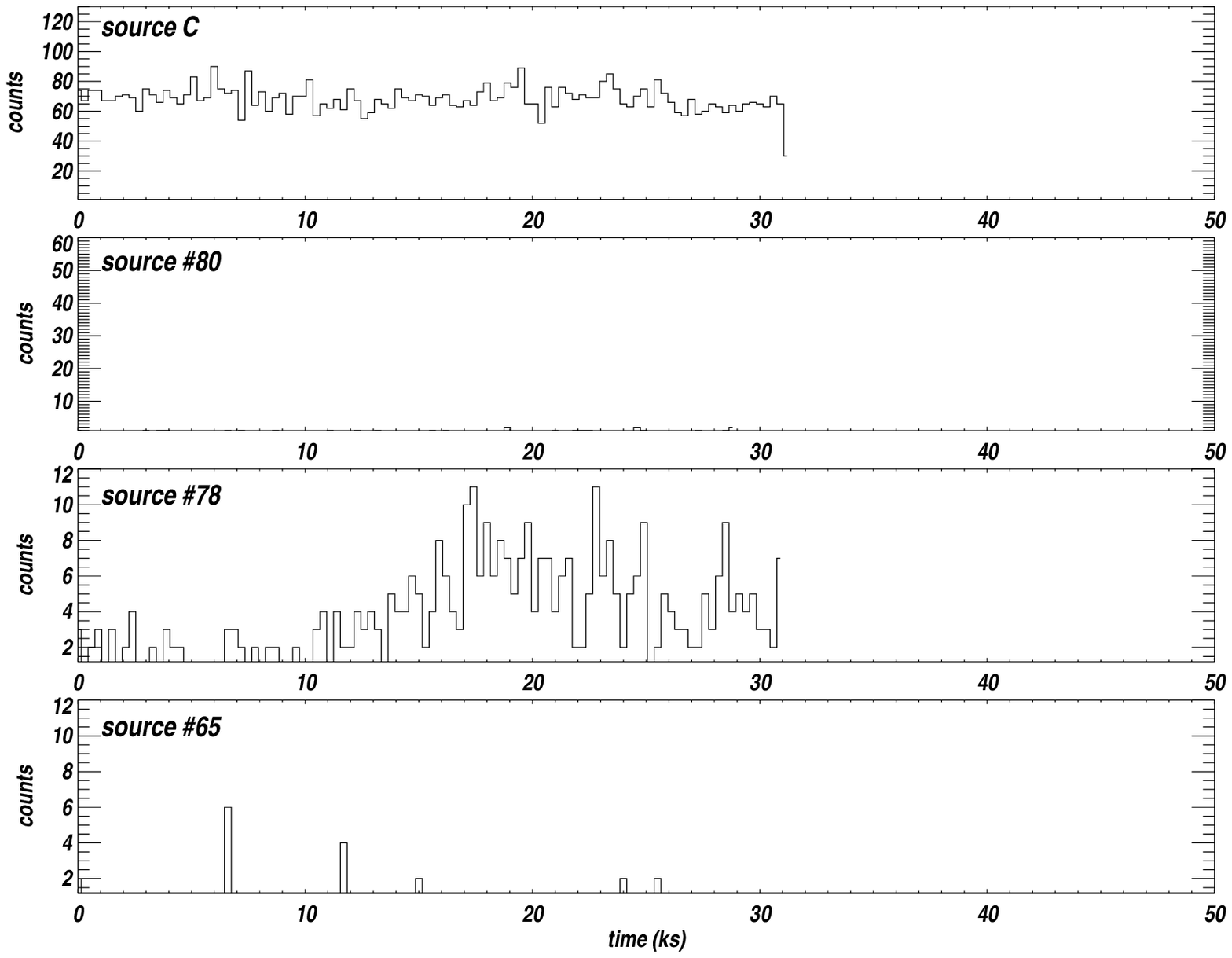}
}
\caption{X-ray light curves of 4 Trapezium sources for the October (left) and the November (right) observa
tions.}
\label{lcurves}
\end{figure*}

\begin{figure*}
\centerline{\epsfxsize=17.0cm\epsfbox{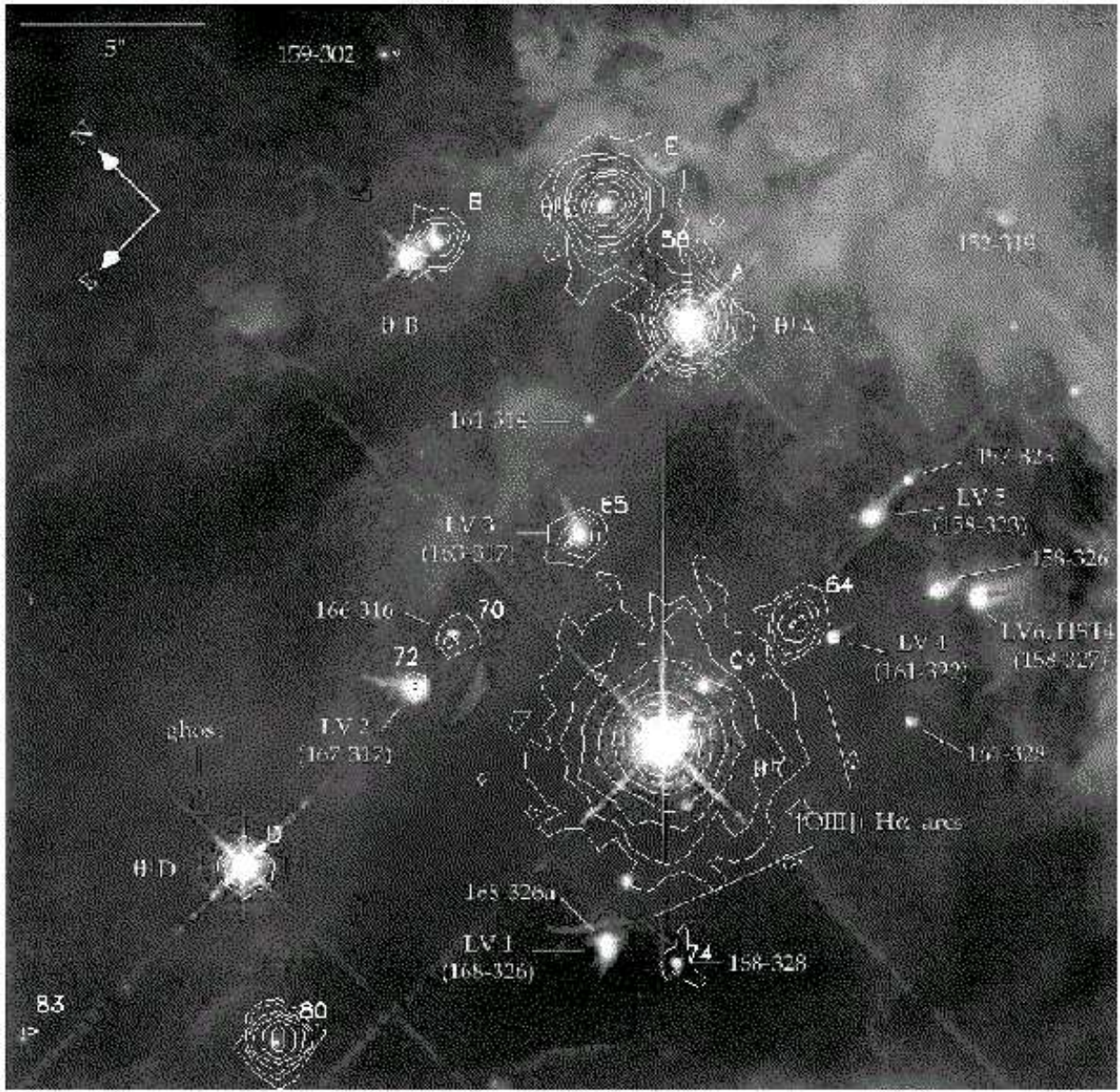}}
\caption{X-ray contours of the merged Chanda events overlaid onto an optical HST PC observation (from
Bally et al. 1998). The field of view is 30"x30" and the contours represent total
counts. The lowest contour
represents a flux at about the detection threshold.
\label{field}}
\end{figure*}

\end{document}